\documentclass[sort&compress
    ,final            
  ]
  {aipproc}

\layoutstyle{8x11double}

\begin{document}

\title{Reforming a large lecture modern physics course for engineering majors using a PER-based design}


\author{S. B. McKagan}{
  address={JILA, University of Colorado, Boulder, CO, 80309, USA}
}
\author{K. K. Perkins}{
  address={Department of Physics, University of Colorado, Boulder, CO, 80309, USA}
}
\author{C. E. Wieman}{
  address={JILA, University of Colorado, Boulder, CO, 80309, USA},
  altaddress={Department of Physics, University of Colorado, Boulder, CO, 80309, USA}
}

\begin{abstract}
We have reformed a large lecture modern physics course for engineering majors by radically changing both the content and the learning techniques implemented in lecture and homework.  Traditionally this course has been taught in a manner similar to the equivalent course for physics majors, focusing on mathematical solutions of abstract problems.  Based on interviews with physics and engineering professors, we developed a syllabus and learning goals focused on content that was more useful to our actual student population: engineering majors.  The content of this course emphasized reasoning development, model building, and connections to real world applications.  In addition we implemented a variety of PER-based learning techniques, including peer instruction, collaborative homework sessions, and interactive simulations.  We have assessed the effectiveness of reforms in this course using pre/post surveys on both content and beliefs.  We have found significant improvements in both content knowledge and beliefs compared with the same course before implementing these reforms and a corresponding course for physics majors.
\end{abstract}

\maketitle


\section{Introduction}
It is well-documented that PER-based interactive engagement techniques improve student learning in introductory physics courses.  However, the use and study of these techniques has been much less common in upper-level physics courses.  Some physics professors who accept the use of interactive engagement techniques in introductory physics courses claim that these techniques are inappropriate for more advanced courses.  At the University of Colorado, we are working to systematically reform both introductory and advanced courses in physics and other sciences, and to document the results of these reforms.  In the fall of 2005 and spring of 2006, the authors taught and reformed Physics 2130, a modern physics course for engineering majors \cite{2130}.  This course is the third semester of physics, and is typically taken by sophomore mechanical engineering majors and senior electrical engineering majors.  This is the last physics course that most of these students will take.

This course has been reformed using a research-based design, based on the following principles learned from Physics Education Research:
\begin{enumerate}
    \item Using interactive engagement techniques can lead to higher learning gains than traditional lecture. \cite{Hake1998a}
    \item Directly addressing common misconceptions can lead to higher learning gains. \cite{McDermott2001a}
    \item Unless physics content is presented in a way that explicitly addresses student beliefs about science, these beliefs tend to become more novice-like. \cite{Redish1998a,Perkins2005a}
    \item People have a limited short term memory, so material should be presented in a manner that reduces cognitive load by focusing on the important points, having a coherent structure, and eliminating nonessential details. \cite{Mayer2003a}
    \item For students to gain a conceptual understanding of the material, all aspects of the course, including homework and exams, must address conceptual understanding, not just numerical problem-solving.
\end{enumerate}

\section{Process of Reform}
In order to develop a clear set of learning goals for the course, we interviewed seven physics faculty members about the most important concepts they thought students should learn from this course.  These interviews elucidated an important issue in reforming more advanced courses: unlike introductory physics, in which there is a well-established set of topics on which most experts agree, there is no general consensus about what should be taught in more advanced classes.  This issue is particularly acute in this course, which by default has often closely resembled the corresponding course for physics majors, who will see the topics in several later courses.

To determine how to make our course most relevant to our target audience, we met with a group of engineering professors, to whom we posed the question, ``What do your students need to know about modern physics?''  The general consensus was that engineering students need to know about applications of quantum mechanics such as electron devices, lasers, STMs, and MRIs; they need to know about the quantum origin of molecular bonding and material structure; and they need some experience solving differential equations describing physical systems.  The engineering professors said that their students do \textit{not} need to know about special relativity or a lot of abstract mathematical formalism, topics that had typically been emphasized in this course.

To address principle 2, we reviewed the existing PER literature on student difficulties in modern physics and quantum mechanics.  In addition, one of us (SBM) hosted a weekly problem-solving session for students in the course the semester before our reform.  Field notes from this session provided insights into common problems for our student population.

Concurrently with our reform effort, we have been developing the Quantum Mechanics Conceptual Survey (QMCS), a multiple choice survey designed to test student understanding of the fundamental concepts of quantum mechanics \cite{QMCS}.  The questions in the QMCS are based on faculty interviews, studies of textbooks and syllabi, existing conceptual tests of quantum mechanics \cite{Cataloglu2002a,Singh2001a}, research studies of student misconceptions \cite{Ambrose1999a,Styer1996a,Fletcher2004a}, informal observations of students in problem-solving sessions and class, and formal interviews with students.  The student interviews conducted to validate this survey were also useful in gaining a better understanding of student misconceptions.

The content of the course was chosen to reflect the concepts most commonly cited in faculty interviews (fundamental principles of quantum mechanics), the priorities of the engineering faculty (real world applications and the relationship of microscopic principles to macroscopic properties of materials), and expert beliefs about the relevance and coherence of physics (real world applications, grounding in experiment, conceptual understanding, and reasoning development).

\section{Interactive Engagement Techniques}
We encouraged interactive engagement in class by assigning students to 3 person consensus groups for peer instruction.  We asked an average of about 5 questions per 50 minute class, to which students submitted their answers using clickers.  Most of the questions included a period of group discussion.  We used several different kinds of clicker questions, including interactive lecture demos in which we asked students to predict the results of an experiment, usually demonstrated with a simulation; eliciting misconceptions in order to address them; polling students to find out more about their background or what they wanted us to address; asking students to work through difficult multi-step problems; and quizzes on the assigned reading.  During clicker questions, the three instructors as well as three undergraduate learning assistants circulated through the room in order to facilitate group discussions and to listen and report back on what students were thinking.  We occasionally used other interactive engagement techniques in class, for example working through a tutorial on quantum tunneling.

Outside of class, we encouraged interactive engagement by hosting collaborative problem solving sessions where students could work together on homework.  These sessions were staffed by instructors, undergraduate learning assistants, and graduate teaching assistants, all of whom were trained to facilitate discussion and help students work to figure out the answers on their own, rather than telling the students the answers.  These sessions were voluntary, but attendance was encouraged by advertising their value and making the homework sufficiently difficult that students could seldom complete it on their own.  According to the end of term survey, one third of the students attended the problem-solving sessions at least 80\% of the time, and another third attended 20-60\% of the time.

We have developed a suite of interactive computer simulations
on quantum mechanics specifically for this course as part of the Physics Education Technology (PhET) Project \cite{phet}.  These simulations follow research-based design principles and are extensively tested through student interviews and classroom studies.  In the course, we incorporated simulations into lecture, clicker questions, and homework.  The homework included a large number of guided inquiry activities designed to help students explore and learn from the simulations.  By providing visual representations of abstract concepts and microscopic processes that cannot be directly observed, these simulations help students to build mental models of phenomena that are often difficult to understand.

The simulations incorporate many of the principles listed in the introduction, such as reducing cognitive load by focusing student attention only on essential features.  For example, many students have difficulty understanding the circuit diagram for the variable voltage supply in the photoelectric effect experiment, which distracts them from seeing the main point of the experiment.  By illustrating the variable voltage supply as a battery with a slider, our Photoelectric Effect simulation eliminates this distraction.

\section{Conceptual Understanding and Reasoning Development}
Throughout the course, we focused on helping students develop conceptual understanding and reasoning skills, such as making inferences from observations and understanding why we believe the ideas of quantum mechanics.  This was emphasized in all aspects of the course.  We wrote all our own homework, which was online and composed of computer-graded multiple choice and numeric questions, and TA-graded essay questions.  The homework was designed to be extremely difficult conceptually, though only moderately difficult mathematically.  Thus students were required to write essays explaining a conceptual model or to determine the underlying reasons for a complex physical phenomenon.

For example, students worked through a series of homework questions using the Lasers simulation to build up an understanding of how a laser works, at the end of which they had to write essays on questions such as why a population inversion is necessary to build a laser and why this requires atoms with three levels instead of two.

\section{Real World Applications}
We incorporated applications into every aspect of the course, presenting at least one application of each major concept discussed.  We presented photomultiplier tubes as an application of the photoelectric effect; discharge lamps, fluorescent lights, and lasers as applications of atomic structure and transitions, alpha decay and STMs and applications of quantum tunneling, LEDs and CCDs as applications of the quantum theory of conductivity, and MRIs as an application of spin.  A lecture on Bose Einstein Condensation tied together many of the concepts introduced throughout the course.

\section{Textbook}
Finding a textbook appropriate for this course was difficult, given the focus on conceptual understanding and applications, which are not suitably covered in standard texts.  The first semester we used Tipler and Llewellyn \cite{Tipler2002a}, a popular modern physics textbook that was consistent with our level of math and contained most topics we covered.  Students complained about the text on a weekly basis, both verbally and in feedback forms, and our top students reported that they stopped reading the textbook because they couldn't understand it.  Many students used our power point lecture notes, which were posted online, as an alternative to the textbook.  The second semester we switched to portions of volumes 3 and 5 of Knight's introductory physics textbook based on physics education research \cite{Knight2004a}.  This textbook is at a lower level than our course mathematically, and it does not include many of our topics such as the time-independent Schrodinger equation.  However, the pedagogical focus for the topics it does include is much more consistent with our with our approach.  There were almost no complaints about the textbook second semester, and the average student ranking of the usefulness of the textbook for their learning on a scale of 1 (not useful) to 5 (a great deal) went up from 2.1 to 3.2.  The usefulness rankings for other aspects of the course did not change significantly between the two semesters, and were between 3.5 and 4.3, with the posted lecture notes receiving the highest ranking.

\section{Assessment of Course}
We used several methods for assessing the effectiveness of instruction in this course, including giving the QMCS pre/post as measure of conceptual learning and the Colorado Learning and Attitudes about Science Survey (CLASS) \cite{Adams2006a} as a measure of the change in student beliefs about science.  We have also done several studies to assess learning in particular content areas of the course, the results of which will be reported elsewhere.

We gave the QMCS as a pretest and posttest during the two semesters in which we taught this course (Fa05 and Sp06) both to our students (engineering majors) and to the students in the corresponding course for physics majors.  We also gave it as posttest to both the engineering and physics majors' courses the semester before our reforms (Sp05).  We use the other four modern physics as baseline ``traditional'' courses \footnote{Two of these courses used clickers, but in a quite limited fashion.}.  It is worth noting that our class size was approximately 180 students both semesters, more than double the typical size of this course in previous semesters, about 80.  The physics majors' course ranges in size from about 30 to 80.

Table \ref{QMCSresults} shows the QMCS results.  We calculate the average normalized gain (<g>) for each course, which measures how much students learned as a fraction of how much they could have learned.\footnote{<g> and its uncertainty $\Delta$<g> are computed as in Ref. \cite{Hake1998a}.}  The values of <g> for the reformed courses are consistent with typical normalized gains on the Force Concept Inventory (FCI) in reformed introductory physics courses \cite{Hake1998a}.  There are wide variations in <g> for the traditional courses, but all are consistently lower than in the reformed courses.  It is interesting to note that the physics majors started consistently higher than the engineering majors, but ended up lower than the engineering majors in the reformed courses.

\begin{table}
\begin{tabular}{lccccc}
\hline
{\bf Course} &  {\bf Pre} & {\bf Post} &  {\bf <g>} & {\bf $\Delta$<g>} &    {\bf N} \\
\hline
Ref. Eng. Sp06  &         30 &         65 &       0.49 &       0.04 &        156 \\

Ref. Eng. Fa05  &         32 &         69 &       0.54 &       0.05 &        162 \\

Trad. Eng. Sp05  &         (30) &         51 &       0.30 &       0.05 &         68 \\

Trad. Phys. Sp06  &         44 &         64 &       0.37 &       0.15 &         23 \\

Trad. Phys. Fa05  &         40 &         52 &       0.21 &       0.06 &         54 \\

Trad. Phys. Sp05  &         (44) &         63 &       0.34 &       0.08 &         64 \\
\hline
\end{tabular}
\caption{Average percent correct responses and normalized gains on 12 common questions on QMCS for six modern physics courses.  We do not have pretest data for Spring 2005, so in the analysis of these courses, we have assumed that the average pretest scores would have been the same as the following spring.  Because the survey is still under development, different versions were used different semesters.  This analysis includes only the 12 common questions that were asked all three semesters.}
\label{QMCSresults}
\end{table}

It should be noted that the QMCS covers only the fundamental concepts of quantum mechanics, and not any of the applications that constituted a substantial fraction of our course.  However, all six courses spent a comparable amount of time on the material covered by the QMCS, since the engineering majors' course in Sp05 covered statistical mechanics and the physics majors' courses covered special relativity, neither of which were covered in the reformed courses or tested in the QMCS.

It is difficult to evaluate the relative success of our treatment of the real world applications that constituted a major part of our course, because this material is simply not covered in other courses.  However, this is likely to impact students beliefs about science, and this can be compared with other classes.

We gave the CLASS to assess student beliefs.  It is a well known result \cite{Redish1998a,Perkins2005a} that in a typical physics course, these beliefs tend to shift towards novice-like.  In other words, students leave most physics courses believing that physics is less coherent, less logical, and less relevant to their everyday lives than when they started the course.  There is some evidence that, because the subject is so abstract and counterintuitive,  teaching modern physics can have a negative impact even in courses where special efforts are taken to address beliefs  \cite{McCaskey2005a}.

Table \ref{CLASSresults} shows that while the traditional modern physics courses had large shifts towards novice-like beliefs, there were no statistically significant shifts in the overall beliefs of students in the reformed courses.  While it is difficult to pinpoint a single cause of this difference, it seems reasonable that the emphasis on real world applications and reasoning development helped students to see the subject as more relevant and coherent.

\begin{table}
\begin{tabular}{lccccc}
\hline
{\bf Course} &  {\bf Pre} & {\bf Post} & {\bf shift} & {\bf $\Delta$shift} &    {\bf N} \\
\hline
Ref. Eng. Sp06  &       66.1 &       67.1 &        +1.0 &        1.1 &        135 \\

Ref. Eng. Fa05  &       70.2 &       68.0 &       -2.1 &        1.1 &        150 \\

Trad. Eng. Sp05  &       68.5 &       60.5 &       -8.0 &        1.9 &         55 \\

Trad. Phys. Sp06  &       72.1 &       67.2 &       -4.9 &        2.4 &         25 \\

Trad. Phys. Fa05  &       78.6 &       72.9 &       -5.7 &        1.5 &         47 \\

Trad. Phys. Sp05  &       78.5 &       74.8 &       -3.7 &        1.5 &         61 \\
\hline
\end{tabular}
\caption{Average percent favorable (expert-like) responses on CLASS for the same six modern physics courses shown in Table \ref{QMCSresults}.  The shifts for the reformed courses are not statistically significant, unlike the traditional courses, which all have large statistically significant shifts down.}
\label{CLASSresults}
\end{table}

\section{Conclusions and Next Steps}
We have reformed a large lecture modern physics course for engineering majors by implementing peer instruction, collaborative homework sessions, and interactive simulations, and by emphasizing real world applications, conceptual understanding, and reasoning development.  These reforms have been successful in producing increased learning gains and eliminating the substantial decline in student beliefs.  We are now working on the next step, archiving and sustaining these reforms.  This course will be taught next semester by a different professor in the PER group, who will use our materials.  We will continue to work to improve the course and package it in a way that is easy to pass on to other instructors.

\begin{theacknowledgments}
  We would like to thank the undergraduate learning assistants and graduate teaching assistants who helped with the course, as well as the PhET team and the Physics Education Research Group at the University of Colorado.  This work was supported by the NSF.
\end{theacknowledgments}



\bibliographystyle{aipproc}   

\bibliography{PERCproceedings2006}

\IfFileExists{\jobname.bbl}{}
 {\typeout{}
  \typeout{******************************************}
  \typeout{** Please run "bibtex \jobname'" to optain}
  \typeout{** the bibliography and then re-run LaTeX}
  \typeout{** twice to fix the references!}
  \typeout{******************************************}
  \typeout{}
 }

\end{document}